# Deep learning guided image-based droplet sorting for on-demand selection and analysis of single cells and 3D cell cultures

Vasileios Anagnostidis,[1] Benjamin Sherlock,[1] Jeremy Metz,[1] Philip Mair,[2] Florian Hollfelder,[2] and Fabrice Gielen [1]

[1] University of Exeter, Living Systems Institute, Exeter, United Kingdom
[2] University of Cambridge, Department of Biochemistry, Cambridge, United Kingdom

**Abstract**

Uncovering the heterogeneity of cellular populations and multicellular constructs is a long-standing goal in fields ranging from antimicrobial resistance to cancer research. Emerging technology platforms such as droplet microfluidics hold the promise to decipher such heterogeneities at ultra-high-throughput. However, there is a lack of methods able to rapidly identify and isolate single cells or 3D cell cultures. Here we demonstrate that deep neural networks can accurately classify single droplet images in real-time based on the presence and number of micro-objects including single mammalian cells and multicellular spheroids. This approach also enables the identification of specific objects within mixtures of objects of different types and sizes. The training sets for the neural networks consisted of a few hundred images manually picked and augmented to up to thousands of images per training class. Training required less than 10 minutes using a single GPU, and yielded accuracies of over 90% for single mammalian cell identification. Crucially, the same model could be used to classify different types of objects such as polystyrene spheres, polyacrylamide beads and MCF-7 cells. We applied the developed method to the selection of 3D cell cultures generated with Hek293FT cells encapsulated in agarose gel beads, highlighting the potential of the technology for the selection of objects with a high diversity of visual appearances. The real-time sorting of single droplets was in-line with droplet generation and occurred at rates up to 40 per second independently of image size up to 480 x 480 pixels. The presented microfluidic device also enabled storage of sorted droplets to allow for downstream analyses.

## 1. Introduction

Deciphering population heterogeneities is a long-standing goal in cellular biology. At the level of single cells, such heterogeneities are usually observed at the genomic, transcriptomic or phenotypic levels[1]. Likewise, populations of multicellular constructs display heterogeneities for every single cell of the constructs as well as overall organisational and morphological variations that have proven useful to mimic the complexity of organs or tumours[2]. Strikingly, even clonal spheroids originating from single cells can lead to very diverse gene expression profiles[3]. We can dissect these non-uniformities by developing technologies that can isolate single cells or 3D cell cultures, preferably at higher throughput to allow population analysis with statistical significance. Compartmentalisation of cells in water-in-oil microdroplets[4], or hydrogel beads[5, 6], is one promising avenue to cellular analyses. Droplet microfluidics enable a vast number of biological applications from single-cell -omics studies, ultrahigh-throughput molecular screens and high-throughput screening of three dimensional cell cultures[7-11]. A key advantage of droplet microfluidics lies in the possibility to apply a selection pressure and isolate rare phenotypes at high-throughput. Two areas of research that would particularly benefit from further development of droplet sorting technologies are single cell micro-encapsulation and screening of 3D cell cultures.

Indeed, micro-encapsulation of discrete objects is governed by Poisson statistics, resulting in many droplets being empty or containing more than one object of each type. Targeting an average occupancy of one cell per droplet results in only 36% of droplets with a single cell, reducing the actual biological throughput 3 fold below the physical one. In addition, cell doublets confound results of experiments where homogenous single cell populations are required for correct data interpretation. This can occur when cells stick to each other or are co-isolated and can lead to potentially crucial cells being missed out. For instance, Macosko et al. found that cell doublet rates ranged from 0.36% to 11.3% for relevant cell concentrations[12]. The likelihood of multiple occupancy can be reduced by heavily diluting the cell solution, but this acts to lower biological throughput, increasing reagent costs and the risk of cell death[13]. In applications where object mixtures in precise numbers are required, Poisson distributions become a significant limitation to the biological throughput. For instance, the probability of having two different single objects at target occupancy of one per droplet is only 13%. This is highly relevant because recently developed single-cell methods require increasingly complex mixtures of objects to be combined in a droplet. For instance, analysis of the genome and transcriptome of thousands of cells simultaneously can be done by co-encapsulation of one bead carrying DNA as an identifier together with single cells to analyse[9, 14]. Other applications have leveraged the throughput afforded by microfluidic experimentation platforms to co-encapsulate cells with functional binders[15]. The One-Bead-One-Compound (OBOC) approach to drug screening requires one DNA-encoded drug bearing bead to be co-encapsulated with a target cell[16]. Antibody-producing cells can be screened by co-encapsulation with beads capturing released antibodies or directly with antigen-producing cells[13, 17].

Research in three-dimensional (3D) cell cultures also benefits from the throughput afforded by microdroplet encapsulation which can be used as a template for hydrogel bead formation in which cells proliferate[6]. Every bead represents a unique microniche where cells can form physiologically relevant models to investigate cancer cell behaviour, tissue engineering from stem cells and high-throughput drug testing. [18-21] However, assay reproducibility issues, especially for quantifying drug efficacy, have plagued the field and led to numerous inconclusive studies[22-24]. There is a proven need to pre-select and study 3D cell cultures with uniform sizes (i.e. size, morphology, levels of development) so that drug screen or phenotypic assays can be reliably assessed across homogeneous samples.[25] Yet, high-throughput microfluidic sorting of specific subpopulations of micro-encapsulated 3D cell cultures has not been attempted to date. This is partly because many studies focus on spheroid compaction (aggregation of a large number of cells) during short timescales (1-2 days) instead of spheroid growth over several days/weeks, where phenotypic variations will likely be more significant .[26]

## 1.1 On-demand sorting of single cells and 3D cell cultures

High-throughput sorting has long been performed by flow cytometers which can also accommodate large objects such as 3D cell cultures. However, flow cytometers fail to identify objects with non-uniform shapes and morphologies, because they mostly rely on light scattering properties. More modern imaging cytometers can acquire high-throughput imaging data but do not yet allow for sorting based on image analysis.[27]

In contrast, microfluidic engineering can be combined with bespoke optical setups to achieve droplet selection based on many types of readout. Droplet selection has historically relied on fluorescence read-outs but has since been expanded to absorbance, fluorescence lifetime, image analysis, or Raman scattering.[28-30] Image-based approaches are becoming increasingly popular, in part due to increases in image processing algorithms, interfacing speed, and better processing hardware with multiple studies in recent years[31-33]. Bright-field imaging enables real-time visual inspection of droplets and can be coupled with droplet sorting[34]. In contrast to single-point measurement approaches relying on light fluctuations imaging provides crucial morphological information, which can be used for instance to differentiate objects of similar size.

### 1.2 Deep learning for object counting and morphology screens

Object recognition from bright-field images can be achieved using rules-based or machine learning strategies. Rules-based algorithms such as Hough transforms can recognize objects of specific shape but cannot easily adapt to changing appearances so that they require extensive manual recalibration for every experimental condition. In contrast, machine learning algorithms learn the rules from labelled data and enable accurate and robust classification of images.[35,36] They remove inherent biases linked to arbitrarily deciding sets of rules to recognize objects. Machine learning techniques are particularly well suited to droplet classification because ultra-large annotated training sets can be obtained rapidly. The objective function to identify objects randomly located in a droplet is also well defined. Girault et al. implemented template-matching algorithms can identify parts of images that match a pre-defined template.[32] However, they are likely not to perform well when object appearance deviates from the template. Amongst other machine learning techniques, convolutional neural networks have the intrinsic property to learn high and low-level features and are being implemented in a vast number of fields, from industrial machine vision to medical diagnostics.[37] This feature learning ability is especially useful to learn complex, specific morphological attributes and can be used in conjunction to object counting. We provide a comparison table (ESI §1) to highlight the unique advantages of CNN-based methods for droplet sorting.

In this article we report a deep learning guided image-based real-time sorting method for on-demand selection of micro-objects in microfluidic devices. We train deep neural networks to classify droplets according to their content as they flow and triage them based on the type and number of objects they contain, using three classes defined as containing zero, exactly one and more than one object. We demonstrate complex selections by applying the same algorithms to recognize single cells co-encapsulated with other objects, paving the way towards on-demand multiple Poisson distribution. Finally, we show selection of 3D cell cultures and isolate the ones displaying successful proliferation from a large initial seed population of individual cells.

## 2. Experimental

### 2.1 Chip fabrication and materials

Microfluidic device fabrication was done using classical soft lithography techniques. Devices made up of two layers were produced by sequential spin-coating with SU-8 and UV cross-linking before a final wet development step. The first layer included a flow-focussing structure (height, 80 µm; width, 120 µm at the junction) consisting of two separate inlets for aqueous phases, a mixing channel, an additional inlet for oil addition, a sorting junction and a serpentine loop. The second layer with only a serpentine trapping channel was developed for droplet deceleration and trapping (height, 150 µm; width, 150 µm). Design files will be deposited on DropBase (openwetware.org/wiki/DropBase). The carrier oil phase was prepared using fluorinated oil HFE-7500 (Fluorochem Ltd) containing 1% (w/v) 008-Fluorosurfactant (RAN Biotechnologies, Inc.) for in-line sorting and 0.5% (w/v) for agarose generation. Polyacrylamide beads (PA) were purchased from Droplet Genomics (Ø 65 µm) and polystyrene spheres (Ø 10 µm) from Polysciences.

### 2.2 Cell culture and harvesting

The breast adenocarcinoma MCF-7 human cell line (Sigma) and human embryonic kidney (HEK) cell line 293FT (Thermo Fisher) were cultured in Dulbecco's Modified Eagle Medium (DMEM, Gibco) supplemented with 10% Fetal Bovine Serum (Gibco) and 1% of Glutamax (Gibco) at 37 ℃ with 5% $CO_2$. The medium was filtered through a filter (0.2 µm; Sartorius) before usage. Cells were passaged when reaching approximately 80% confluency using 0.05% trypsin-EDTA (1X, Gibco) to detach the cells , before centrifugation at 1200 rpm for 5 minutes and re-suspension in fresh medium. For single cell experiments, the cells were counted, placed in medium containing 15% (v/v) OptiPrep density gradient medium (Sigma) and passed through a 20 µm cell strainer (Falcon, Fisher Scientific) before experiments to minimize cellular aggregates. For multicellular spheroid formation, 293FT cells were harvested, diluted to a density of 10^6/mL, and mixed with liquid agarose (ultra-low gelling temperature, Sigma) dissolved in PBS (1X, Gibco), resulting in a final gel concentration of 1% w/v. The cells were subsequently encapsulated using a flow-focussing device (height, 80 µm; width, 80 µm at the junction). Flow rates used were 25 µL $min^{-1}$ and 5 µL $min^{-1}$ for the carrier oil phase and for the cells, respectively, resulting in gel bead generation at 160 Hz. The collected emulsion was then cooled at 4 °C for 15 minutes to allow gelification. Demulsification was done using an antistatic gun[38], after which the gel beads were centrifuged at 300 rpm for 3 minutes. Supernatant was removed, replaced by 1 mL of fresh culture medium and the beads were culture for 4 days to enable cell proliferation. We then collected and concentrated all the beads by centrifugation at 300 rpm for 3 minutes before resuspension in 100 µL of fresh medium. The beads were subsequently loaded in a PTE tubing attached to a glass syringe (SGE air-tight, 1 mL) and connected to an inlet of the sorter. The microfluidic sorter produced microdroplets in which single gel beads were encapsulated. The sorted and waste/control droplets containing the gel beads were de-emulsified using the same protocol for bead generation and transferred to a 24-well plate (Greiner BioOne). The beads were imaged with a High-Content Screener (ImageXpress, Molecular Devices).

### 2.3 Computing

We used a Windows 7, 64-bit operating system with an Intel i5-6500 3.2 GHz processor with 32 GB RAM and CUDA capable dedicated graphics card (GeForce 1080 GTX Zotac). Software protocols were written in Python 3.6 using Tensorflow 1.3.0 and OpenCV 3.3 libraries. Images were saved in real-time on a local SSD drive. Scripts are provided as Supplementary Information.

### 2.4 Optics

The setup is based around an inverted microscope (Brunel, SP-98-I)in which the camera port of the trinocular was adapted to host the imaging, triggering and high-speed video acquisition (Fig. 1B). A white LED light source (CoolLED, pe-100) was for its output stability. Optical parts and connectors were purchased from Thorlabs. Exact dimensions are provided in Fig. S1†.

### 2.5 Image acquisition

A Field-Programmable Gate Array (FPGA) was programmed to read the input voltage of a photodetector (PDA36A), recording light fluctuations in the field of view and output a 5V trigger signal to the fast area scan camera (Pike F032B) whenever the signal exceeded an arbitrary, manually set threshold. We fixed light intensity so that incoming power on the microfluidic chip was constant (~1.5 mW). The exposure time and gain of the area scan camera were fixed to 100 μs and 15 dB respectively. For the training and testing set, images of 480 x 480 pixels were grabbed as soon as available in the camera's RAM and immediately saved onto an SSD drive, cropped down to a circular region-of-interest fully enclosing a droplet of diameter 150 μm, and normalized (mean subtraction followed by division with the standard deviation). For the test images, the convolutional neural network (CNN) model applied to a single image at a time. The output of the CNN gave a predicted class together with an associated probability which was used for droplet selection.

### 2.6 Droplet formation and in-line sorting

The aqueous and carrier oil phases were loaded into 1 mL glass syringes (SGE) and 5 mL plastic syringes (BD Plastipak; sterile needles, 25G), respectively, and pushed by syringe pumps (Nemesys, Cetoni). Polythene tubing was purchased from Scientific Laboratory Supplies (0.38 mm ID x 1.09 mm OD). The flow rates for oil and aqueous solutions were 10 μL $min^{-1}$ and 1 μL $min^{-1}$, respectively. The oil used for spacing out droplets was flown at 15 μL $min^{-1}$.

### 2.7 Droplet selection

The TTL trigger signal to activate the electrodes was generated by National Instruments data acquisition (DAQ) board USB-6009 and the trigger pulse was directed towards a function generator which in turn produced a 10 kHz signal for 15 ms (10 kHz, square wave, 6 Vpp, TG2000, TTi instruments). This signal was amplified 100 times by a high-voltage amplifier (Trek 6001) and the electric field carried through saturated salt water solution was propagated within ad-hoc channels in the microfluidic device.[39]

## 3. Results

### 3.1 A monolithic microfluidic device for droplet generation, mixing, sorting and storage.

The same microfluidic device produced droplets for in-line imaging and sorted them before storing some of them. Droplets had an average diameter of 150 µm and were generated at rates between 10 and 40 per second depending on the flow rates used (Fig. 1A-1). Following generation, the content of droplets was mixed during travel through a serpentine shaped channel (Fig. 1A-2). This mixing of separate solution eliminated inner interface phases (e.g. seen with cell suspension+ 15% OptiPrep mixed with PBS buffer) and improved object clarity for the real-time imaging. The droplet-to-droplet interval was subsequently increased to facilitate sorting by injecting more carrier oil just before a sorting junction (Fig. 1A-3). This increased motion blur during snap-shots by fewer than 2 µm at maximum speed. Real-time imaging was performed approximately 450 µm downstream to the central post of the sorting junction. Two separate channels loaded with a concentrated salt solution, and located adjacent to the sorting junction were responsible for inducing a dielectrophoretic force, required for sorting, on the droplet.[39, 40] The sorting channel then led to a storage line having a high depth so that droplets decelerated and approximately 30 of the last sorted droplets could be stored in the sequence they were sorted (Fig. 1A-4). Hydrostatic pressure induced by the outlet tubings was balanced at the start of each experiment by adjusting outlet tubing heights while monitoring droplet trajectories at the sorting junction. Similar to a previous study,[34] the imaging setup relies on acquiring images of flowing droplets rather than continuously processing video data. This configuration enables stable droplet imaging at a fixed, pre-determined position and prevents the processing of superfluous inter-droplet frames. We designed the setup so that droplet imaging and visualisation of the droplet sorting process could be done using the same microscope objective (Fig. 1B). Using a beam-splitter and relaying focusing lenses placed along the imaging path, we were able to achieve dual field-of-view imaging. The larger area (800 µm x 800 µm) was enough to image the sorting Y-junction and confirm successful droplet selection while the smaller area (160 µm x 160 µm) enabled imaging of single droplets. A photodetector (PD) was used to detect light intensity fluctuations due to droplet interfaces and an FPGA used the signal from the PD in real-time to send a trigger pulse initiating droplet image acquisition. An iris was placed before the photodetector to modulate the integrated area being measured and used to maximize signal-to-noise which facilitates reproducible triggering.

### 3.2 Application to the classification of 3 types of objects

We used polyacrylamide hydrogel beads (PA, Ø 65 µm), polystyrene spheres (PS, Ø 10 µm) and MCF-7 cells (MCF7, Ø 15-20 µm) to test and validate the platform. The refractive index difference between the PA beads and water, despite the high water content of the PA hydrogels, was sufficiently high to observe a faint contrast at the PA/water interface. We further increased that contrast using an iris diaphragm (aperture diameter 1 mm) placed after the collimated white LED light source which increased the depth of field (depicted as I1 in Fig. 1B). This resulted in a slight thickening of the droplet interface which did not hamper object identification. PS spheres have a high refractive index (1.58) and appeared typically as black and white dots occupying 25 x 25 pixels at camera level, while the MCF-7 cells had a vast number of possible appearances and covered typically a 35 x 35 pixels area. Although the cells and beads were much smaller than the channel height (80 µm), they could be identified away from the central focusing plane thanks to the large depth-of-field.

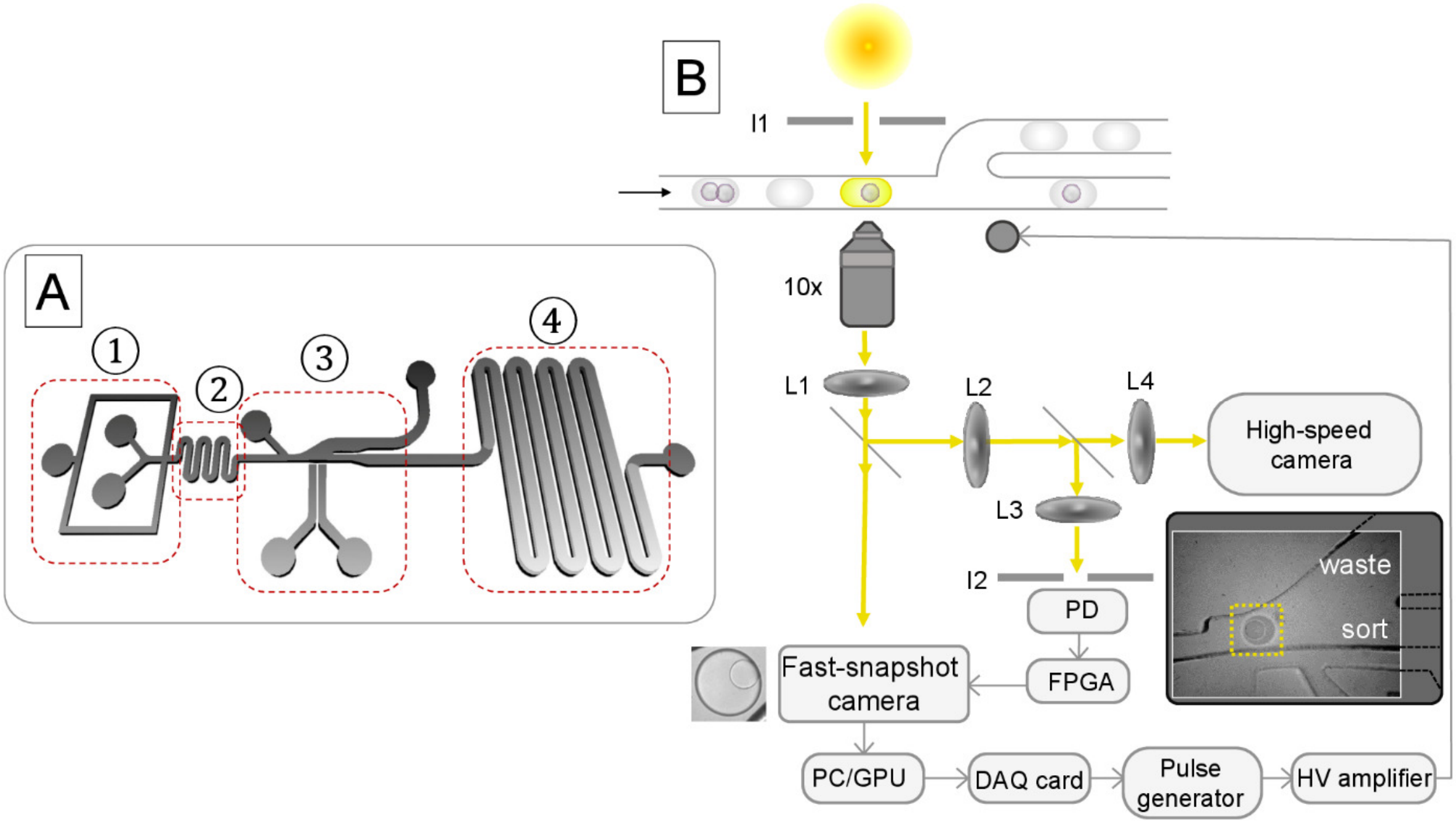

**Fig. 1** (A) The droplet sorting device integrates droplet generation combining two different solutions (1), mixing (2), respacing and sorting (3) followed by in-line storing (4). (B) Schematic of the optical setup for high-speed imaging and recording of droplet sorting events. A microfluidic device is imaged with an inverted microscope with transmitted white light passing through an iris (I1). This light is split three ways: towards a fast area scan camera, a photodetector (PD) and a high-speed camera. Four focussing lenses were placed along the imaging path so that the high-speed camera could image a large field of view (800 µm x 800 µm) while the firewire camera imaged only a portion of it (approx. 4 times smaller: 160 µm x 160 µm). The signal from the PD was used by an FPGA to trigger the acquisition of a single frame by the scan camera. A second iris (I2) was mounted to adjust the light intensity detected by the PD. A Python script was used to classify any newly acquired image and send in response a triggering pulse, generated by a data acquisition card, towards a high-voltage amplifier deflecting a droplet via dielectrophoresis. Plano-convex lenses had focal lenses of 50 mm for L1 and L2, 25.4 mm for L3 and L4.

Images could be acquired and saved to an SSD drive at tens to hundreds of images per second using a Firewire 800 interface depending on image size (Fig. S3A†). Acquisition (grabbing) alone took less than 1 ms but saving time increased exponentially with pixel number and was typically 22 ms for 480 x 480 pixel images.

### 3.3-Setup of the neural networks and time performance

Convolutional neural networks (CNNs) require labelled training data upon which models are trained. Such image training sets must encompass the largest variability in appearance to train a model that is robust to experimental variations (e.g. which can be used not only intraday but also interday). Consequently, when acquiring initial data sets, we deliberately changed focus during acquisition to simulate differences in focal planes. After an initial acquisition of typically a few thousand images (the number of training images required is examined in Fig. 3A), we manually sorted them into three classes. For all the CNN models presented in this study, we used three classes representing 'empty', 'single object' and 'multiple objects' droplets. We chose such three classes to showcase the potential of CNNs for object counting. However, two classes could also be used albeit with slightly decreased performance (i.e. with 'single object' and 'non single object' classes) (Fig. S2).We used a classical CNN architecture (Fig. 2) made up of 3 convolutional layers, a dense layer with 128 neurons and a 40% dropout layer before the final fully connected layer with 3 classes. The same kernel size of 15 x 15 was used for all layers. We used 10 epochs with a fixed learning rate of $10^{-3}$.The time it took for training using a single GPU was typically 7 minutes (Fig. S3B†) for 480x 480 pixels images with 1200 images per class and 10 epochs (each image was passed 10 times through the network). After training, the CNN was applied in real-time on single images being received by the area scan camera. With the GPU, it took 5 ms to process an image independently of image size (Fig. S3C†) up to 480 x 480 pixels. Different CNN architectures were tested in relation to the training and testing time which shows that increasing the number of filters and kernel size leads to worse timing performance (Fig. S4†). Using 478 x 478 pixels and parameters shown in Fig. 2, we could reach a maximum sorting throughput of 40 Hz. Under standard conditions, system latencies were negligible and did not interfere with the sorting process. A time limit of 12 ms per image was applied by precaution to prevent erroneous sorting.

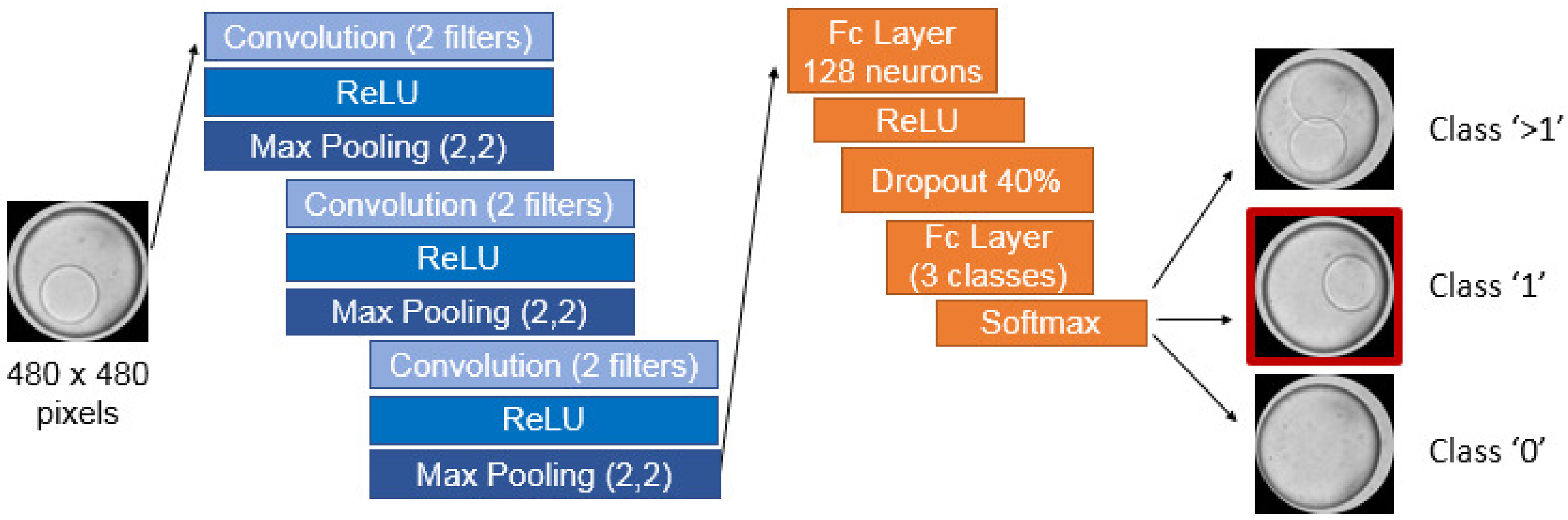

**Fig. 2** Deep neural network architecture. An image is classified as belonging to one of three classes (no object, one object and >1 object). Kernel sizes for the three layers were 15 x 15.

### 3.4 Optimization of the CNN architecture

We optimized the network hyperparameters by analysing accuracy on model validation sets for the number of layers, kernel size, and number of filters per layer using a fixed [Convolution-Rectified Linear Unit (ReLu)-Max Pooling] motif. We repeated this optimization for PA, PS and cells using training images amounting to 300 images for PA beads, and 300 images rotated 10 times (3,300 in total) for PS and MCF-7 cells. The resulting accuracies and losses for a model validation made up of 20 images per class set are displayed in Fig. S5-S6†.

Although the performance of the models seems to follow similar trends for the three objects tested, the main difference is seen in the reproducibility and losses associated. For instance, models generated to recognize cells had generally higher losses than PA beads. Examples of dependence upon many individual parameters of the CNNs are provided in Fig. S7†, S8† and S9†. In the end, we selected the simplest architecture that minimized testing time (i.e. with minimum number of layers, filters and kernel size) and maximized accuracy (as shown in Fig. 2). However, there are a multitude of other network implementations that lead to efficient classification.

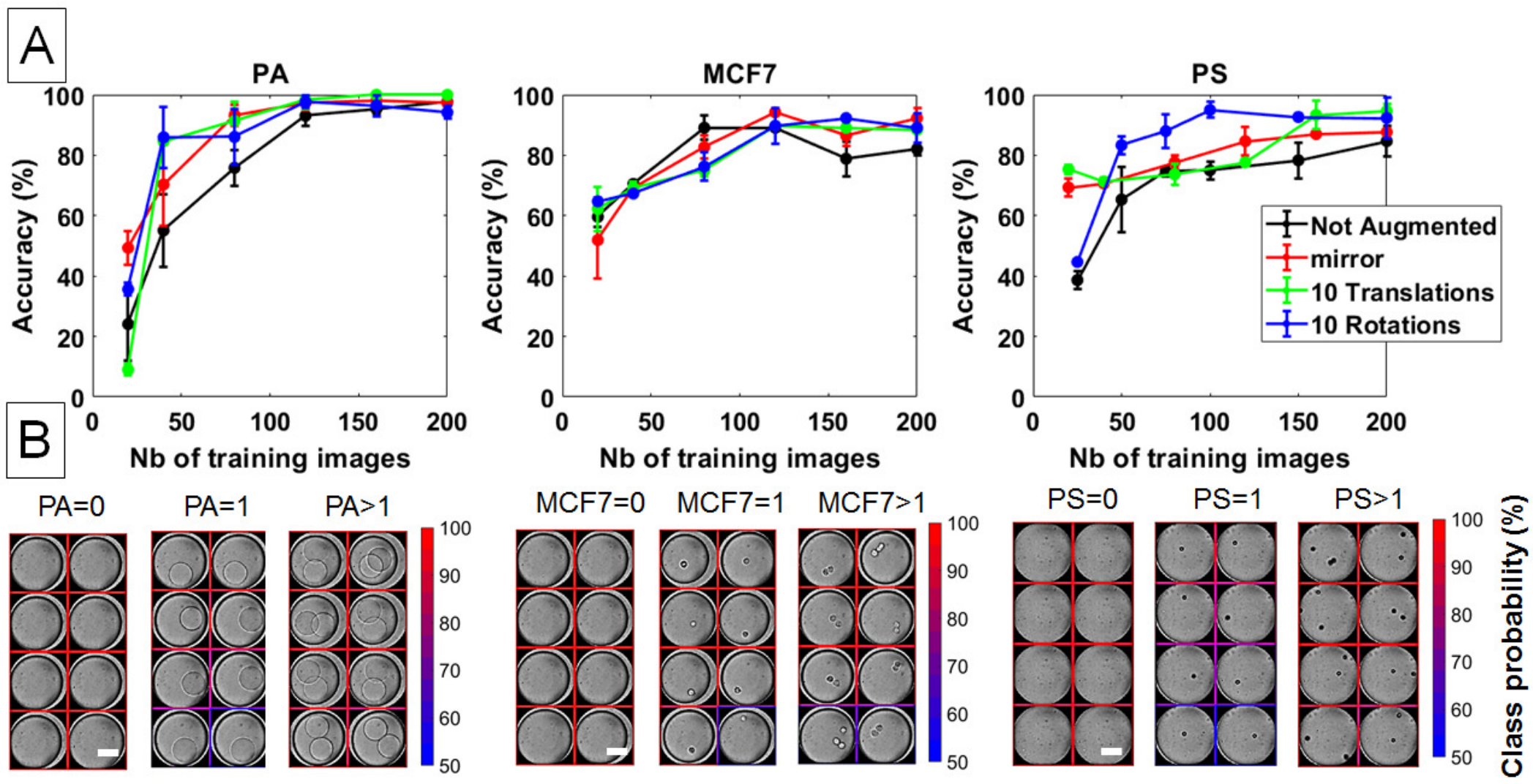

**Fig. 3** (A) Relationship between the number of training images and the efficiency of the image classification algorithms for PA, PS and MCF7. Data augmentation can be used to improve accuracy. The transformations tested were: 10 regular rotations per image, 10 random translations per image (+-20 pixels) (n= 3). 'Mirror' refers to horizontal, vertical and horizontal + vertical flip per image. (B) Examples of correctly classified images after passing through trained CNN models for PA, MCF7 and PS. Associated probabilities are displayed with a colour frame around every image. Scalebars= 65 μm.

### 3.5 Number of training examples

A crucial parameter is the number of images that are required for building a good model (Fig. 3A). To test its importance, we picked an equal number of images per class (200) preventing class imbalance. The validation sets were made up of at least 15 images per class. For objects with a consistent appearance such as PA, the model accuracy increased rapidly and exceeded 90% when given only 125 images per class. The various appearances of cells and PS spheres complicated their correct identification and therefore required more training data.[30] For the cells, the accuracy did not increase continuously with number but had an overall trend towards higher accuracies at higher number. The accuracy for PS spheres quickly increased to above 80% with diminishing returns afterwards. Overall, training with only a few hundred pictures resulted in satisfactory performance and alleviated the tedious task of manually labelling scores of training examples.[31]

Increasing the size of the training pool of images artificially, also known as data augmentation, is a common method for improving a CNN performance without having to provide additional data. Typical transformations include rotations, flipping, translations or noise addition which are added to the original images and form an augmented training dataset. Such transformations would for instance mimic cells located at different locations within a droplet and correct models that take light illumination directionality into account. We found that rotations, translations or mirroring can improve accuracy (Fig. 3A and SI §7).
Fig. 3B displays examples of correctly classified images for the 3 objects tested with optimized neural networks. The filters from activation layers of the models, especially from the second ReLU layer, revealed correct geometrical feature detection: edges of large circles for PA, hot spots for the PS beads and textured spots for the cells (Fig. S11-S13†)(activation layers are used to gain insights about what features of the images are learnt by a particular layer and visualize a CNN internal decision structure).

### 3.6 Robustness of models and drift correction

In this study, we fixed exposure parameters to avoid altering the appearance of objects, in particular to avoid blurring effects induced by the motion of objects during long exposure times. However, it is important to know whether models obtained with training in certain imaging conditions can be used in others. We tested CNN models robustness by assessing the classification of PA beads using various exposure times and gains (Fig. S14 and S15). We found that there were no losses in classification performance as long as we operated below camera saturation levels. We ascribe this robustness to the normalization of the images before processing by the CNN. Other imaging conditions can be altered such as light collimation, type of microscope objective, channel depth, or droplet size. Such experimental constraints will likely restrict the CNN models we generated to our specific platform. However, our study sets the scene for how to replicate such results.

When the models did not immediately classify with satisfactory accuracy, additional images taken on the spot were added to the initial training library for drift correction. The newly trained networks were usually markedly better and permitted good sorting. The time to apply the correction was below 30 minutes for image addition (around 100 per class) and network retraining.

### 3.7 Classification with object mixtures: towards multiple Poisson selections

Specificity of a model for a given object is important to prevent misclassification induced by the presence of other objects in the same droplet. Therefore, we trained models that can anticipate the presence of visually different objects by recognizing single cells in the presence of one or multiple other objects of a second type (here PA or PS). The three mutually exclusive output classes were therefore 'no cell', 'one cell' and 'more than one cell' with any number of other objects (here either PA or PS). We have imaged mixtures of MCF-7 cells with PA or PS and evaluated the accuracy in identifying single cells in a mixture. To this end, we used training sets with ~600 images for MCF7 alone, ~1800 for MCF7 in the presence of PA and ~1200 for MCF7 in the presence of PS.

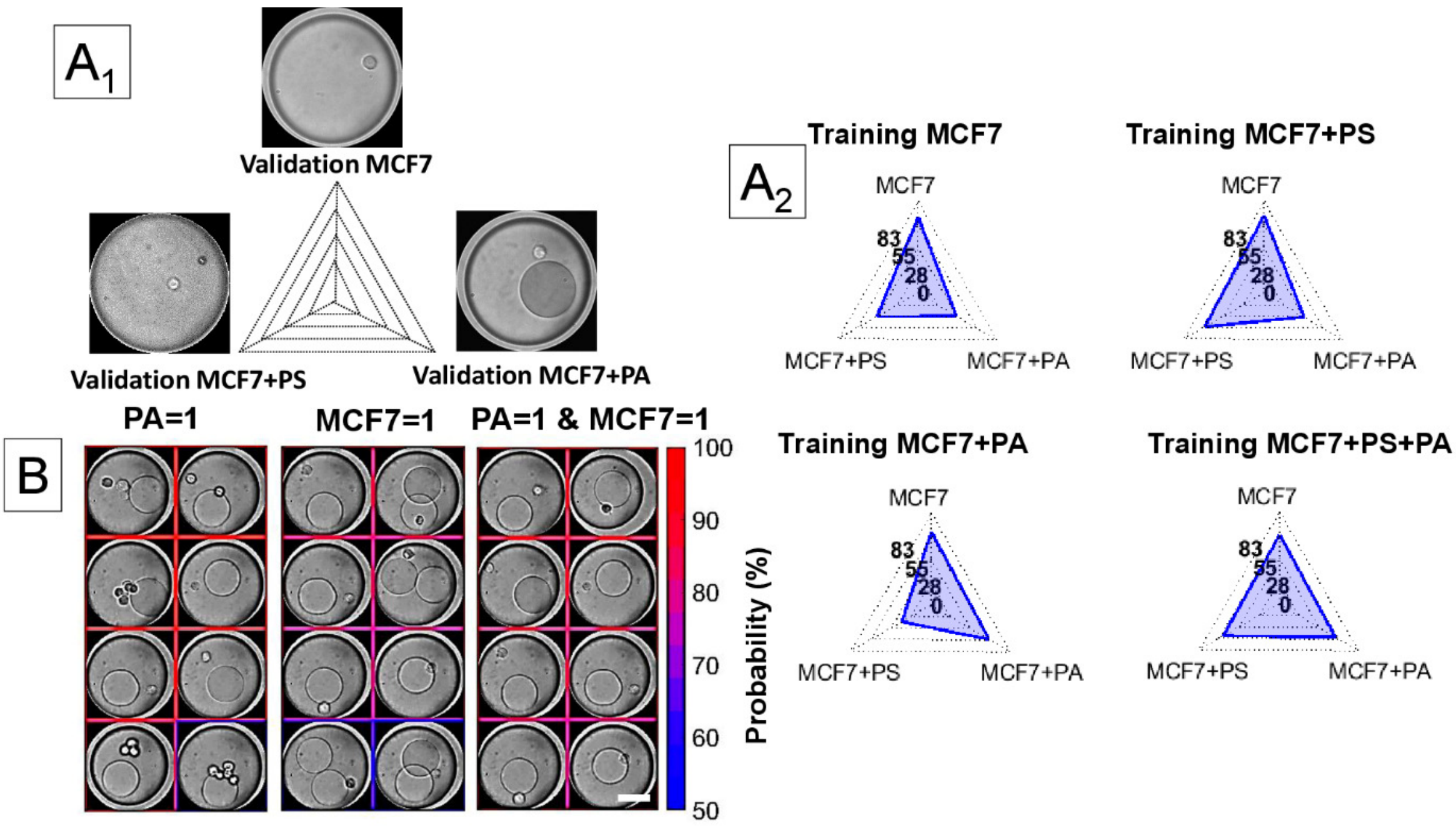

**Fig. 4** ($A_1$) Influence of the composition of the training set on classification accuracy when validating with images comprising mixed objects. Example images for each of the validations are displayed. ($A_2$) Accuracies obtained by training for classification of images with MCF7 cells shown in $A_1$ with training sets comprising MCF7 alone, MCF7 and PS, MCF7 and PA and all objects. Images with MCF7 cells can be classified correctly only if the network has been trained to anticipate the presence of other objects. (B) Examples of single objects correctly identified in a mixture of mixed objects for single PA mixed with MCF7 (left), single MCF7 cells with PA (middle) and, using both models, single MCF7 with single PA (right). Scalebar= 65 µm.

Training for mixed objects also included the training set acquired for MCF7 alone. Fig. 4A shows that accuracy increased dramatically when training included examples of objects given in the validation set. When given examples of PS and/or PA in addition to MCF7 alone, the models performed well (80-85% accuracy) on the corresponding validation sets (Fig. 4A, Training MCF7+PS and Training MCF7+PA). When given all types of examples, the models could correctly identify most images whether in the presence of PA or PS (Fig. 4A, Training MCF7+PS+PA). We could use this strategy to identify droplets with two types of single objects: for instance by combining two models, one for PA in the presence of MCF7 and one for MCF7 in the presence of PA, we could identify droplets with exactly one PA and one cell (Fig. 4B).

### 3.8 Validation in real-time sorting experiments

We performed model single object sorting experiments during which we saved all images and counted the number of images misclassified as either false positives (wrongly assigned the 'single object' class for multiple objects) or false negatives (single objects not recognized as such) (Fig. 5). PA models were effective at recognizing single PA beads and therefore false positive/negative rates were very low (<8%). For the cells, typical models led to an overall 20% false positive rate. However, bringing the probability threshold for sorting to 90% led to a decrease down to 10% false positives at the cost of missing out approximately 50% of cells. For PS, the gain for raising the probability threshold was not so high and mainly led to increased false negative rates.

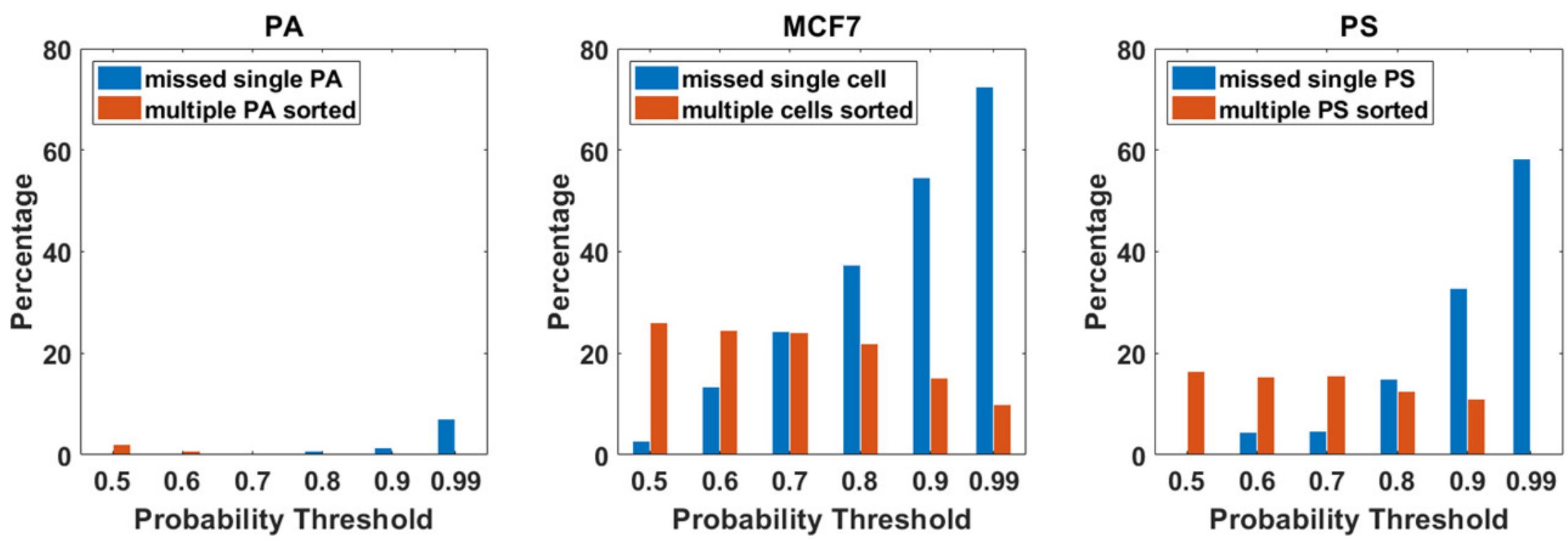

**Fig. 5** Histograms showing the percentage of false positives as a fraction of sorted droplets and false negatives as a fraction of all screened droplets during model sorts. Histograms were built from 141, 451 and 132 images for PA, MCF7 and PS, respectively.

We directed sorted droplets towards a storing line for imaging and validation. The storing serpentine allowed deceleration of the droplets and approximately 30 were stably trapped at the end of the loop. It operated as 'first droplet-in, first droplet-out', similar to a shift-register.[41] Fig.

6(i)-6(iv) displays example images of sorted droplets using PA, MCF7 and PS CNN models. The storing loop kept sorted droplets in sequence so that droplets of varying content could be alternatingly selected, e.g. empty drops and droplets with single PA (Fig. 6, ii). The sorting models used approximately 300 images per class for PA, 100 images per class augmented by rotations (20x, total of 5000 images) for PS and 100 images per class augmented by mirroring (horizontal, vertical and both) and rotations (1x) for MCF7 (total of 4000 images). Finally, we demonstrated identification of single MCF-7 cells mixed with PA beads using a larger training set of 2000 images per class. Fig. 6-v shows the images used for classification by the CNN and the corresponding droplets stored after sorting (Fig. 6-vi). Interestingly, two cells identified as single cells turn out to be doublet (indicating cell division or distinct cells stuck together) but this could not be easily seen on the images used for classification.

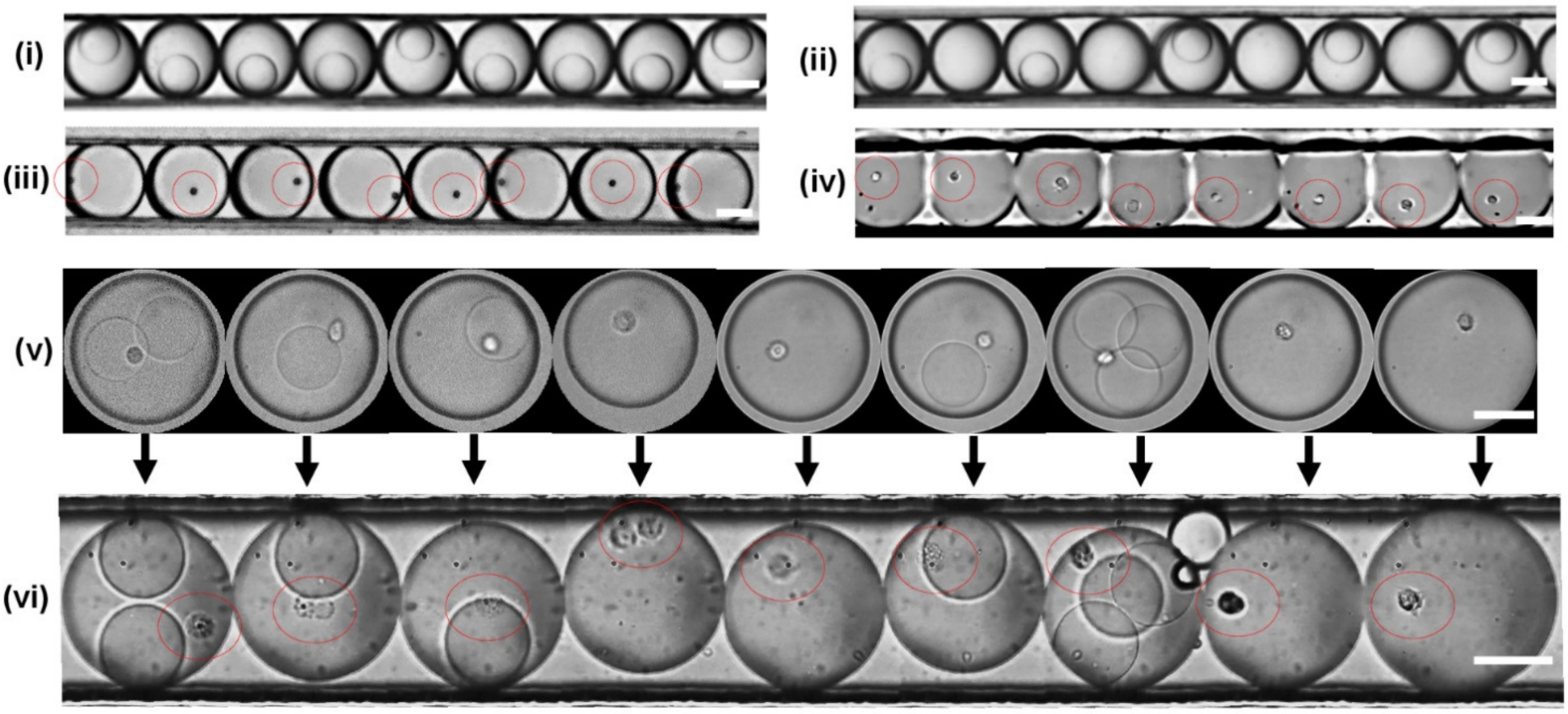

**Fig. 6** Assembled images of stored droplets at the end of the storing line. (i) Single PA beads, (ii) Alternation of empty and single PA beads. (iii) Single PS beads. (iv) Single MCF7 cells. (vi) Single MCF7 cells with PA beads with corresponding images used for sorting (v). Some cells appear as doublets (or dividing cells) but are not obviously visible in the image used for sorting. PS and cells are highlighted by red circles. All scale bars represent 65 µm.

**3.9 Selection of multicellular spheroids**

We demonstrated classification of objects with variable appearance by growing 3D cell cultures within agarose microgels. Here, we used droplet sorting as a tool for spheroid isolation. Approximately 50,000 cultures were generated with an average number of cells per bead of 1.5 cells per bead and incubated in a cell incubator. A large proportion of the cells did not proliferate and eventually died, resulting in a large population of beads either empty or containing dead single cells (80%) after 4 days incubation. We ascribe poor cell viability to a combination of high shear rates during encapsulation, gelation at 4 ℃ and overall long time of manipulation steps outside cell incubators (~45 min). We aimed at using the image-based sorter to concentrate the proportion of grown spheroids from this initial heterogeneous population. We harvested cells in beads after 4 days of growth and re-injected them in the microfluidic sorter. The complete workflow is displayed on Fig. 7A. For training, we manually selected 100 spheroids whose diameters were at least 30 µm. The second and third classes were made up of 100 beads with single cells and 100 droplets either empty or with empty agarose beads, respectively (no augmentation). The same neural network as shown in Fig. 2 was used. Despite the variations in morphologies, the model was able to accurately identify the spheroids (Fig. 7B). Example activation layers (Fig. S16†) show the recognition of combination of textures and edges. After training, we screened approximately 5,500 beads and isolated 250 in 12 minutes. The rate of false negatives and positives was found to be 3% and 12%, respectively, at 90% probability. The sorted were then demulsified and imaged using a high-content screener prior to placing them back in the cell incubator. Comparison images of pre- and post-sorting are displayed in Fig. 7C-7D. We obtained a 4-fold enrichment in formed spheroids, from 20% before sorting to close to 80% after sorting. The mismatch between observed false positive during sorting and after sorting was attributed to spheroids escaping the gel beads during the process of resuspension in culture medium. The spheroid diameters (Fig. 7E) ranged from 30 to 80 µm indicating the CNN model is robust to variation in spheroid size. We confirmed spheroid viability 24 hours after sorting using live/dead stains (Fig. S17†).

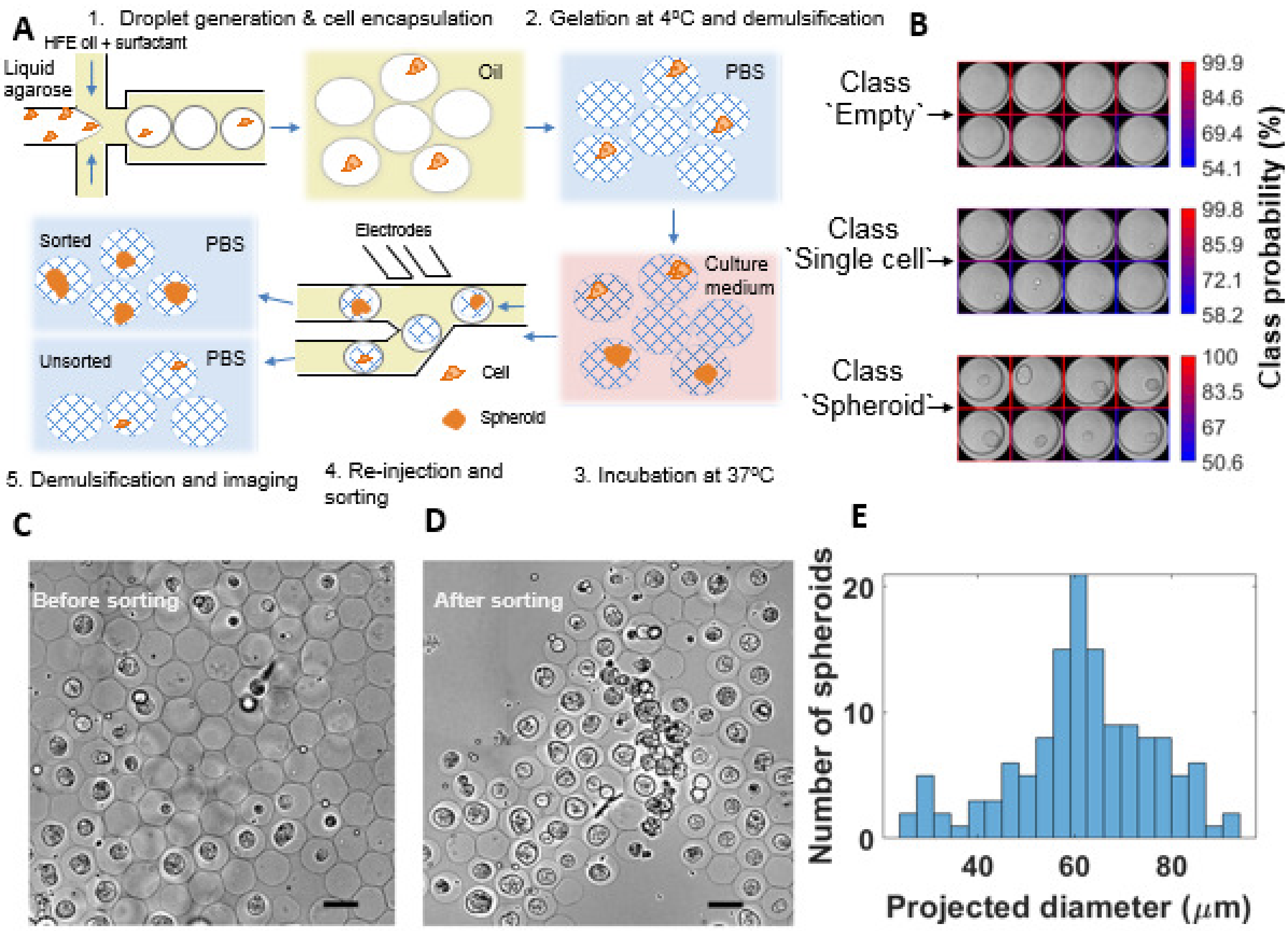

**Fig. 7** Multicellular spheroid selection by CNNs. (A) Flow schematic for droplet generation, cell incubation and sorting. (B) Validation images for CNN training with three classes (empty, single cells and spheroids). (C) Bright-field image of beads before sorting. (D) Bright-field image of beads after sorting. (E) Histogram of projected spheroid diameters of the sorting sample (n= 134). Scalebars= 100 µm.

## 4. Discussion

Deep learning provides a method for object detection without explicitly deciding sets of rules. The specificity of the models built means the technique is robust with respect to the presence of foreign objects (e.g. dust particles) or droplet polydispersity. The only image pre-processing steps we used were cropping and normalization so that the models did not require any expected appearance. In particular, there was no need for water/oil boundary recognition which may hamper recognition of objects located close to the droplet interface. The composition of training classes can be adapted to the type of selection required. For instance, one can establish a class corresponding to the exact target number of objects as we have shown for 'exactly 0' and 'exactly 1', provided enough training data can be acquired and class imbalance avoided. Often, a model can be too specific to object appearance, location and noise, leading to overfitting. This can be alleviated by accumulating larger image collections encompassing the largest variability in appearance with a view to generalizing models. Several recent papers are tackling robustness of the neural networks by implementing novel types of layers, such as the neighborhood similarity layer[42], or techniques for bypassing noise such as multi-resolution imaging.[43] The arbitrary choice of images, especially for the validation sets, often precludes systematic quantitative analysis of model performance. Additionally, the neural network hyperparameter space is too large to pinpoint the best parameters but our manual optimization indicates that the parameters we used lead to efficient models that allow largely correct sorting decisions with minimal need for further enrichment. The intrinsic stochasticity of these algorithms can mean that different outcomes are obtained when retraining with the exact same images and parameters. Therefore, it is important to validate the models and only keep those with the lowest losses. Interestingly, the rate of false positive/negative events can be predicted from validation tests so that the rate of errors in sorting experiments can be anticipated.
Here, all the objects we studied could be found independently of their location in the droplets even though small PS spheres, for example, represent less than 15% of the droplet diameter. This also implies objects located directly below one another in the focal direction may be partially obscured and missed. However, our study highlights the potential for finding objects away from the point of optimal focus and the possibility to screen objects of different sizes simultaneously. Our work also paves the way for selecting multiple single objects co-encapsulated, which we show can be achieved by training two independent models for each object of interest in the presence of the other.

We envision further directions to improve deep learning image classification of microdroplets. In this work, we show that CNNs can be trained with relatively low number of labelled images (~100 per class) for single object classification and with a simple network structure. Further gains in image classification accuracy could be achieved by accumulating more labelled training data. However, this is a tedious, time-intensive task. This will prove even more problematic when attempting to sort based on small or subtle differences in appearance. There is scope to identify improved and object-specific data augmentation methods, for instance using generative techniques such as GANs[44]. Other types of supervised learning could be used to alleviate labelling efforts, such as semi-supervised image classification.[45] Transfer learning can also be used as an effective technique to use a network pre-trained on other types of images although timing may become a significant issue due to the typical high depths of such networks.

Our 3D cell culture selection demonstrates that large differences in visual appearance (spheroids of varying size and morphology) covered by training data leads to efficient classification of grown spheroids and that CNNs can be used for enrichment of objects of specific sizes in multi-size object collections. This also confirmed that our network architecture can be generically applied to the recognition of many types of objects. From an application perspective, concentrating spheroids gives the opportunity to accelerate imaging applications and perform screens on relevant 3D cell cultures subpopulations having similar sizes.

## 5. Conclusions

We have demonstrated how deep learning based image classification can be coupled to droplet selection with high identification accuracy leading to efficient real-time classification that are expected to eventually match or outperform human annotation. These models learn characteristic features and are highly specific to object appearance. Accumulating ever larger datasets will lead to increasing accuracies in classification. Class imbalance can be problematic when dealing with rare objects (e.g. double Poisson distributions) but can be solved, for example by introducing a class weighing scheme.[46] With the ongoing global research effort in neural network analyses, systematic parameterization of the hyperparameter space will help identify global optima for any type of object.[47] Other algorithms for real-time object detection could also be implemented such as region-based CNNs (e.g. Faster R-CNN, SPPnet, etc.) but are likely to be computationally more demanding.[48] In the future, the use of deterministic, application-specific integrated circuits for real-time machine vision analyses will increase timing performance and sorting rates. By storing the sorted droplets, we have shown that the selected objects can be subsequently imaged or used in downstream assays, paving the way for high-throughput sorting followed by high-content phenotyping. Our technique could be combined with multiplexed sorting for each class of the CNN, separating many types of objects into distinct collection tubes.[49]

Neural network-based droplet sorting droplet sorting will enable selections based on any visual feature allowing for complex and subtle selection pressures. For instance,
we showed that the CNN models adapt to diverse sets of visual appearance by selecting 3D cell cultures according to various diameters, shape and appearances. Multicellular spheroid selections pave the way towards rapid generation and study of organoid disease models mimicking tissue. For instance, our sorter could be used to screen environmental conditions such as matrix stiffness, composition, presence of cell-cell interaction cues that best promote cellular division and proliferation with ultra-large number of cell cultures beyond the current reach of high-content microscopy. Using high-resolution imaging, single cell sorting can be based on smaller, more complex morphological features (nucleus size and shape, cytoskeleton structure, lipid droplet content…). Such sorts could be used as diagnostics by isolating a small subset of rare cells for further analyses. The addition of multimode measurements (e.g. 3D, DIC, fluorescence…) will increase information rate and enable molecular screens in addition to morphotyping.

### Conflicts of interest

There are no conflicts to declare.

### Acknowledgements

This research was funded by the Royal Society (RG170120). This work was also generously supported by the Wellcome Trust Institutional Strategic Support Award (204909/Z/16/Z). Philip Mair was supported by an EPSRC CDT studentship [EP/L015889/1]. Florian Hollfelder is an H2020 ERC Advanced Investigator [695669].

### Notes and references